\documentclass[final]{ias2}

\usepackage{graphicx} 
\usepackage{multirow}
\usepackage{array} 

\usepackage{hyperref}

\def\bea {\begin{eqnarray}}
\def\eea {\end{eqnarray}}
\def\be {\begin{equation}}
\def\ee {\end{equation}}
\def\nn {\nonumber}

\begin{document}

\markboth{Extent of sensitivity of single photon}{Somnath De}
\title{Extent of sensitivity of single photon production to parton
distribution functions}

\author[som]{Somnath De}
\email{somvecc@gmail.com}
\address[som]{Variable Energy Cyclotron Centre \\ 1/AF Bidhan Nagar, Kolkata - 700064, India}

\date{\today}

\begin{abstract}
We have studied the production of single isolated prompt photons in high energy proton-proton 
collisions at the RHIC ($\sqrt{s}$= 200 GeV) and the LHC ($\sqrt{s}$= 7 TeV) energies 
within the frame work of perturbative QCD upto next-to leading order of strong coupling ($\alpha_s$). 
We have used five different parameterizations of parton distribution function (PDF) starting from the 
old CTEQ4M to the new CT10 distributions and compared our results
with the recent single prompt photon data from the PHENIX and the CMS collaborations. 
The prompt photon cross-section is found to be described equally well by all the PDF's within
the experimental errors at the RHIC and the LHC energies. The deviation in the single prompt photon yield for different
PDF sets is within $\pm$ 20\% when compared to CTEQ4M, indicates the upper bound of uncertainty in determining the gluon density. 
The diphoton measurement could be a potential candidate to constrain the gluon distribution inside the proton.
\end{abstract}

\keywords{Prompt photon, PDF, pQCD.}
\pacs{13.85.Qk, 25.75.Dw, 14.70.Bh}

\maketitle

\section{\bf{Introduction}}
The production of single prompt photons of large transverse momenta in high energy hadronic collisions constitutes 
unique signal of the interactions between quarks and gluons at short distances.~\cite{owen_rmp} 
The single prompt photon yield is expected to be sensitive to parton distribution function (PDF) in general and to gluon distribution
in particular of the colliding hadron~\cite{phot-Atlas}-~\cite{Vogelsang-1}. It is also considered an essential ingredient to 
quantify the nuclear modification of direct photon production in the relativistic nucleus--nucleus collisions~\cite{arleo_raa}.
The word \textquotedblleft prompt\textquotedblright is used to identify the class of photons 
which do not come from the decay of large momenta hadrons (e.g., $\pi^0$, $\eta$, etc.).
 The basic mechanisms of prompt photon production at the Leading-order (LO) of strong coupling $\alpha_s$ are
 (i) quark--gluon Compton scattering (qg$\rightarrow$ q$\gamma$), (ii) quark--anti-quark annihilation 
(q\={q}$\rightarrow$ g$\gamma$) and (iii) bremsstrahlung radiation from the 
final state parton (q(g)$\rightarrow$ q(g)+$\gamma$)~\cite{owen_rmp}. 
Because of the the point-like coupling, the photons produced through the processes (i) and (ii) are called 
\textquotedblleft direct\textquotedblright photons.
 The direct photons are often calculated using perturbative quantum chromodyanmics (pQCD) 
which successfully explains the production of particles with  energy E $\gg\Lambda_{QCD}$ (where $\Lambda_{QCD}\sim 200$ MeV).
 The photons originated from the process (iii) are called \textquotedblleft fragmentation\textquotedblright photons.
The evaluation of this contribution requires non-perturbative parton to photon fragmentation function as 
an input along with pQCD. However, at this point we must state that 
the distinction is renormalization scheme dependent and has no physical implication beyond LO. A complete Next-to Leading
 Order (NLO) theory is needed to explain the single and double prompt photon data in hadronic collisions~\cite{Vogelsang-2}.

The momentum fraction probed by the prompt photons of transverse momentum $\mathrm{p^{\gamma}_T}$ for 
proton-proton (\textrm{p+p}) collisions 
at mid-rapidity (y= 0) is given by; $x \approx \mathrm {2p^{\gamma}_T}/{\sqrt{s}}$ where $\sqrt{s}$ is the center of mass energy.
The photon production is more sensitive to the gluon distribution at low $\mathrm{p^{\gamma}_T}$ (i.e., small $x$) and 
to the valance quark distribution towards higher $\mathrm{p^{\gamma}_T}$ (i.e., large $x$) for a fixed $\sqrt{s}$. 
Thus one can hope to determine the gluon distribution g($x$,Q$^2$) unambiguously
from the very accurate prompt photon data. 

The present work is aimed to verify the above assumptions. For this purpose,
we have considered the isolated prompt photon production in \textrm{p+p} collisions at the Relativistic Heavy Ion Collider (RHIC) 
($\sqrt{s}$= 200 GeV) and the Large Hadron Collider (LHC) ($\sqrt{s}$= 7 TeV) energies.

\begin{figure}[ph]
\begin{center}
\includegraphics[width= 7.5cm, clip= true]{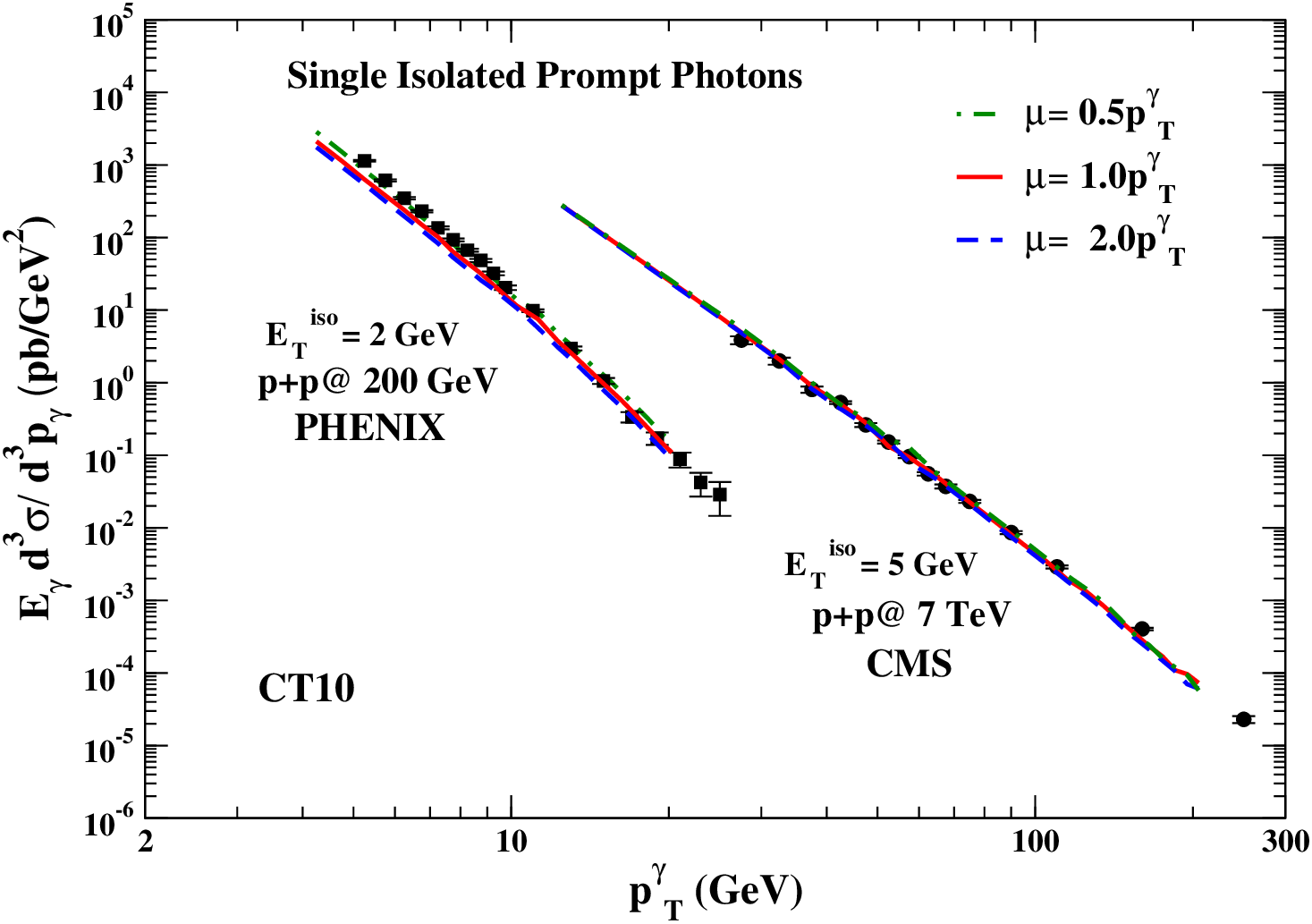}
\end{center}
\caption{(Color online) Scale dependence of single isolated prompt photon production in p+p collisions at the RHIC (200 GeV) and 
LHC (7 TeV) center of mass energies, calculated with CT10 PDF and BFG-I fragmentation function and compared with the measurements
from the PHENIX and the CMS collaborations.}
\label{fig1}
\end{figure}

\subsection{\bf{Theoretical Preliminaries}} 
The isolated prompt photon cross section is found more advantageous than 
the inclusive prompt photon cross section~\cite{Berger,catani}. 
The collider experiments always perform isolated photon measurement to identify the prompt photon signal 
from the large background of secondary decay photons. In addition, the isolation
criterion makes the total photon cross section much less sensitive to the non-perturbative fragmentation process
enabling us to study the small $x$ behaviour of gluon distribution more precisely. 
A photon is said to be isolated if the accompanying hadronic transverse
energy ($\mathrm{E^{had}_T}$) inside the isolation cone is less than some specified value ($\mathrm{E^{iso}_T}$);
\be 
\mathrm{E^{had}_T} \leq \mathrm{E^{iso}_T} \hspace{10 mm} \mathrm{inside \hspace{1.5 mm}R,}
\label{iso}
\ee
where we considered a cone of radius R (=$\sqrt{(y-y_\gamma)^2+(\phi-\phi_\gamma)^2}$) in the rapidity ($y$) and the azimuthal
 angle ($\phi$) space about the photon direction.
It has been shown in Ref.~\cite{catani} that the QCD factorization theorem is valid in all orders of perturbation theory
with the above general isolation criterion (Eq.~\ref{iso}). Hence the isolated prompt photon cross section is a
pertubatively well defined quantity like the inclusive case. In experiment, sometimes the isolation energy is
expressed in terms of a dimensionless parameter $\varepsilon= \mathrm{E^{iso}_T}/\mathrm{p^{\gamma}_T}$.
We write the single isolated prompt photon production cross section as~\cite{Berger,catani}:
\bea
E_{\gamma}\frac{d^3\sigma^{iso}}{d^3p_\gamma}(y_\gamma,p_\gamma, \mathrm{R}) 
= \sum_{i,j} \int \,dx_1  f_{i/A}(x_1,\mu_f^2) \! \int \,dx_2  f_{j/B}(x_2,\mu_f^2) \nn\\
\sum_{c=q,g} \int \, \frac{dz}{z^2} D_{c/\gamma}(z,\mu^2_F) \,
\hat{\sigma}^{iso}_{ij \rightarrow cX} (p_c,x_1,x_2; z_c, \mathrm{R};\mu_f,\mu_R,\mu_F) \Theta(z-z_c) ,
\label{cross}
\eea

where the \textit{f}'s are called the parton distribution functions (PDF's) 
which give the probability for the
incoming partons i and j carrying momentum fractions $x_1$ and $x_2$ inside the hadrons A and B respectively. 
$D_{c/\gamma}(z,\mu^2_F)$ is giving the parton to photon fragmentation probability defined at $z={p_\gamma}/{p_c}$.
When a photon is emitted from \textquotedblleft direct\textquotedblright hard interaction (i.e. $c= \gamma$), 
the photon fragmentation function reduces to $\delta(1-z)$. Following the isolation criterion~(\ref{iso}); 
we find that the fraction $z$ should be $\geq 1/({1+\varepsilon})= z_c$~\cite{catani}.
Thus for small values of $\varepsilon$, the fragmentation process is less contributory to the total
photon cross section compared to the direct process. This is contrary to the inclusive prompt photon production.
The fragmentation function $D_{c/\gamma}(z,\mu^2_F)$ appears in Eq.~\ref{cross} is the same as in case of inclusive photons. 
 $\hat{\sigma}^{iso}$ is the isolated partonic interaction cross section, contains all process upto 
 $\mathcal{O}(\alpha_s^2)$ (for $c= \gamma$) and upto $\mathcal{O}(\alpha_s^3)$ (for $c\neq \gamma$) 
of the reaction $i+j\rightarrow c+X$. 
The dependence of the isolation parameters (R, $\varepsilon$) are fully encoded into the isolated cross section ($\hat{\sigma}^{iso}$). 
$ \mu_f, \mu_R, \mu_F$ are the three different scales associated with the factorization, 
renormalization and fragmentation respectively.

\section{\bf{Results}}
With this theoretical introduction, now we show the results of isolated prompt photon production for \textrm{p+p} 
collisions at the RHIC and LHC center of mass energies (~Fig.\ref{fig1}). Our calculation is based on the Monte
Carlo programme \texttt{JETPHOX}~\cite{jetphox} which evaluates direct and fragmentation photon cross-section 
separately at NLO accuracy. We sum the two contributions to get the total physical cross section at NLO. 
\texttt{JETPHOX} has been found to explain the isolated prompt photon data at the Tevatron energy~\cite{phot-CDF} 
and at the LHC energy~\cite{phot-Atlas,phot-CMS}. 
This programme is very useful in the sense that one can define 
variety of experimental requirements (kinematic, isolation) at the partonic level. Numerous sets of PDF are made 
available in \texttt{JETPHOX} through the LHAPDF library~\cite{Lhapdf}.
We have considered isolation cone of radius R$=$ 0.4 and constant isolation energy cut in these studies. The parameters
of calculations at the RHIC are; $\mathrm{p^{\gamma}_T}$= 4--20 GeV, $|y_\gamma|<$ 0.5, $\mathrm{E^{iso}_T}$= 2 GeV and 
at the LHC are; $\mathrm{p^{\gamma}_T}$= 10--200 GeV, $|y_\gamma|<$ 1.5, $\mathrm{E^{iso}_T}$= 5 GeV. 
We have used the BFG-I parton to photon
fragmentation function by Bourhis \textit{et al.}~\cite{BFG_frag} which includes correction 
in the fragmentation function beyond leading logerthemic approximation. 
The three scales $ \mu_f, \mu_R, \mu_F$ are set equal to a common scale $\mu$, to reduce
theoretical uncertainties in the calculation. The scale $\mu$ further defined as c$\mathrm{p^{\gamma}_T}$ with c= 0.5, 1.0, 2.0.
The scale dependence of the isolated prompt photon production, calculated with CT10 PDF~\cite{ct10} is shown in~Fig.\ref{fig1}.
The results have been plotted against the prompt photon measurement by the PHENIX collaboration at mid-rapidity~\cite{phot-rhic_new} 
and by the CMS collaboration in the rapidity range 0.9 $<\eta <$1.44~\cite{phot-CMS}. 
Our results quite agree with the theoretical curve mentioned in Refs.~\cite{phot-rhic_new,phot-CMS}. 
The scale $\mu= \mathrm{1.0p^{\gamma}_T}$ is found to be a better choice though the relative uncertainty between the scales 
is found to be very small.
\begin{figure}
\begin{center}
\includegraphics[width= 7.2 cm, clip= true]{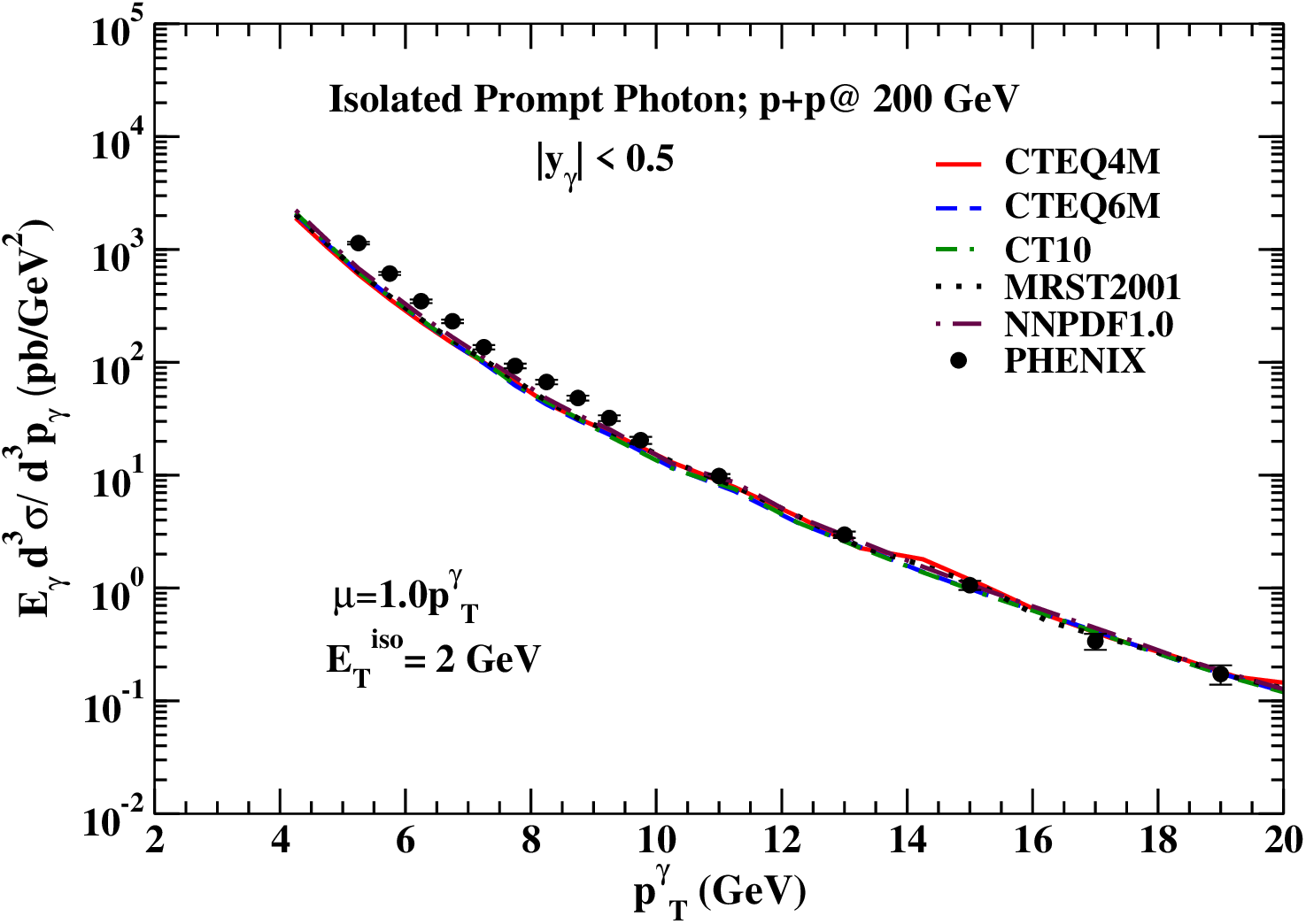}
\includegraphics[width= 7.2 cm, clip= true]{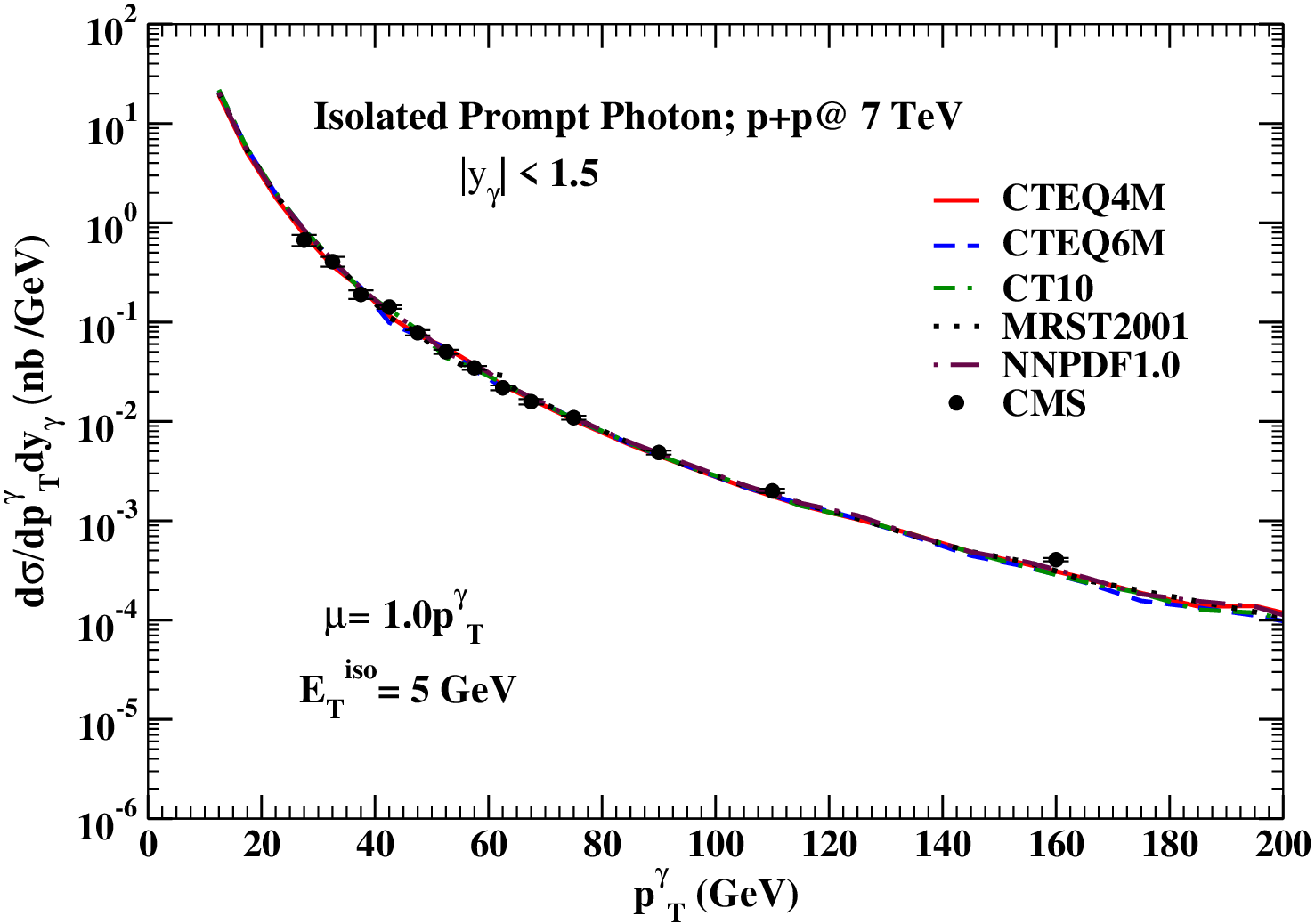}
\end{center}
\caption{(Color online) (Upper panel) Comparison of differential cross section of isolated prompt photons calculated with 
CTEQ4M, CTEQ6M, CT10, MRST2001, NNPDF1.0 PDF, BFG-I fragmentation function and scale  $\mu= \mathrm{1.0p^{\gamma}_T}$ 
for p+p collisions at the RHIC energy. (Lower panel) The same for p+p collisions at the LHC energy.}
\label{fig2}
\end{figure}

Next we have considered four more different parameterizations of PDF namely; 
CTEQ4M~\cite{cteq4}, CTEQ6M~\cite{cteq6}, MRST2001~\cite{mrst1,mrst2} and NNPDF1.0~\cite{nnpdf}.
Our motivation is to check whether the single prompt photon yield is able to distinguish between these PDF sets. We do
hope that the recent prompt photon measurements at the RHIC and LHC experiments can constrain the gluon density in the proton.
The invariant prompt photon yield calculated with the five PDF's at the RHIC and the LHC energies are displayed in~Fig.\ref{fig2}.
We see that starting from very old parameterization like; CTEQ4M to a recent parameterization like; CT10
gives reasonably good description of the data within the statistical errors. A little wiggling observed at different
theoretical curves for $\mathrm{p^{\gamma}_T}>$ 12 GeV at the RHIC and for $\mathrm{p^{\gamma}_T}>$ 120 GeV at the LHC 
can be attributed to the Monte Carlo technique of integration.

\begin{figure}
\begin{center}
\includegraphics[width= 6.2 cm, clip= true]{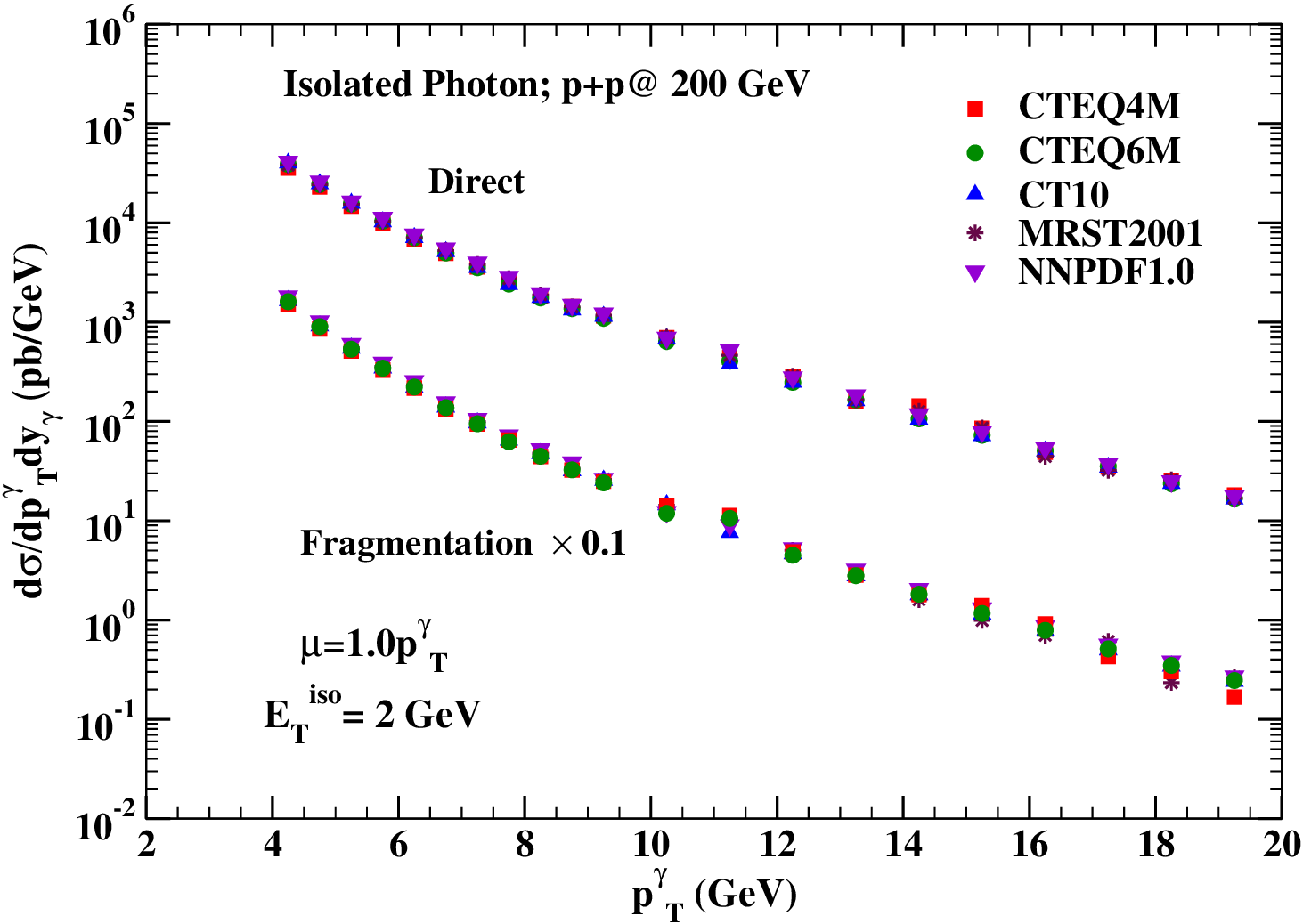}
\includegraphics[width= 6.2 cm, clip= true]{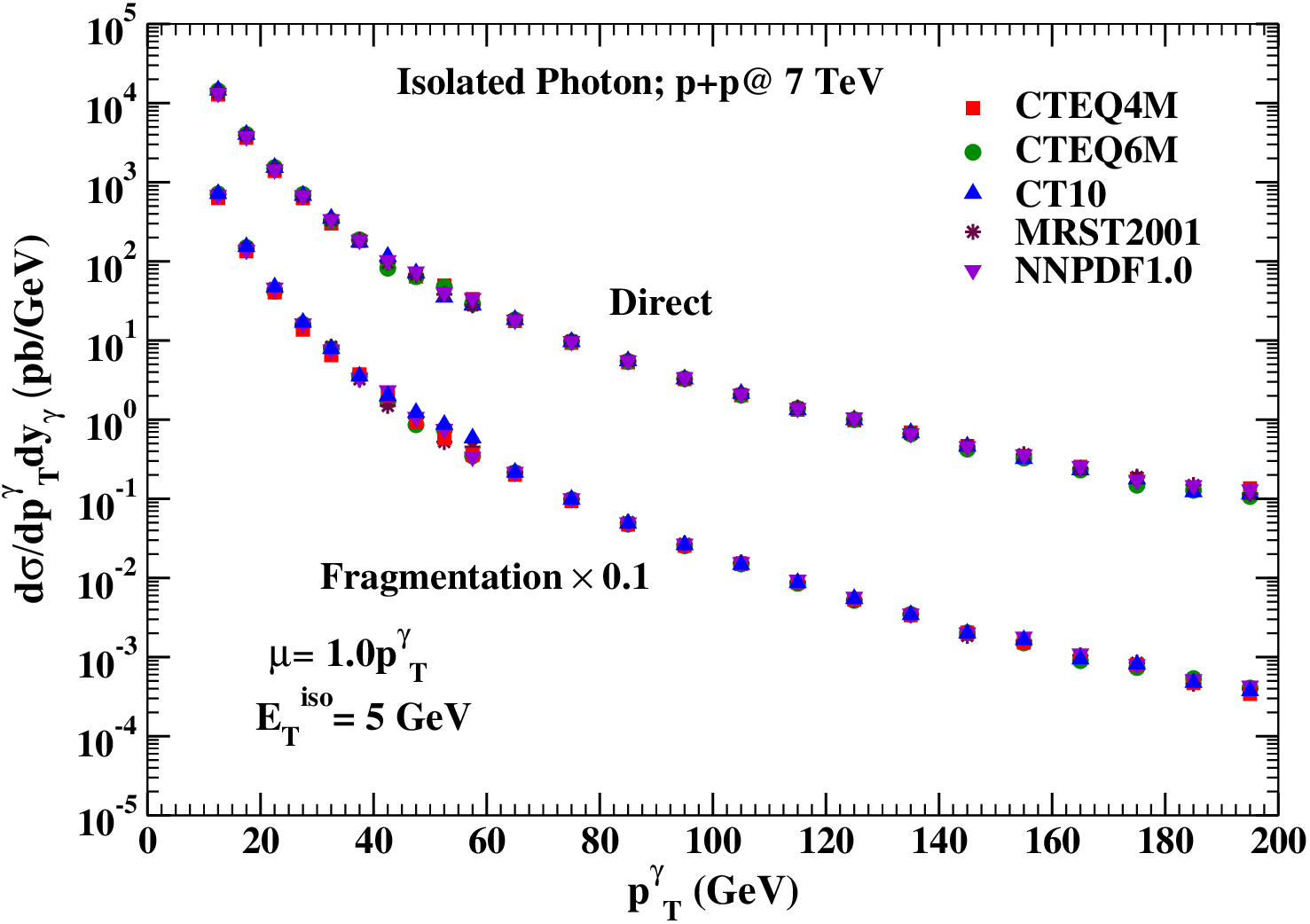}
\end{center}
\caption{\label{fig3} (Color Online) The\textquotedblleft direct\textquotedblright and 
\textquotedblleft fragmentation\textquotedblright contributions to total isolated prompt photon production
in \textrm{p+p} collisions are displayed for the five different PDF's at the RHIC (left) and the LHC energies (right). 
The \textquotedblleft fragmentation\textquotedblright component is scaled by 0.1.}
\end{figure}

The direct and fragmentation contributions to the isolated photon cross section ($d\sigma/dp^{\gamma}_Tdy_\gamma$) 
 at NLO are shown individually in~Fig.\ref{fig3}. 
The NLO corrections to the direct contribution is due to additional real and virtual gluon emission at the LO processes
qg$\rightarrow$ q$\gamma$ and q\={q}$\rightarrow$ g$\gamma$~\cite{dir-Aurenche}. The NLO correction to the fragmentation
is coming from evaluation of generic $\mathcal{O}(\alpha_s^3)$ subprocesses; $i+j\rightarrow c+d+e$ where the parton c 
fragments into a photon~\cite{frag-Aversa,frag-Aurenche}. All these higher oder corrections are included in \texttt{JETPHOX} 
according to the $\overline{MS}$ renormalization and factorization scheme. 
It is known indeed that the total cross section is only the physical quantity at NLO because a direct $\mathcal{O}(\alpha_s^2)$ process 
like; qg$\rightarrow$ qg$\gamma$ contributes the same as LO fragmentation process 
when the fragmentation scale $\mu_F$ is close to 1 GeV~\cite{catani}.
We expect qg (or \={q}g) initiated \textquotedblleft direct\textquotedblright photons could be a good probe 
to track down the varying gluon distribution g($x$,Q$^2$) among the PDF sets.
However as shown in Fig.~\ref{fig3}, we have not found any dependence of the direct and fragmentation contributions to PDF
 at the RHIC and the LHC energies. The fragmentation contribution shown in Fig.~\ref{fig3}, is scaled down by a factor of 
10 for better visibility. We find the direct component has dominant contribution ($\sim$ 80--90\%) 
to the total photon cross section at the RHIC and the LHC energies.

\begin{figure}
\begin{center}
\includegraphics[width= 6.2 cm, clip= true]{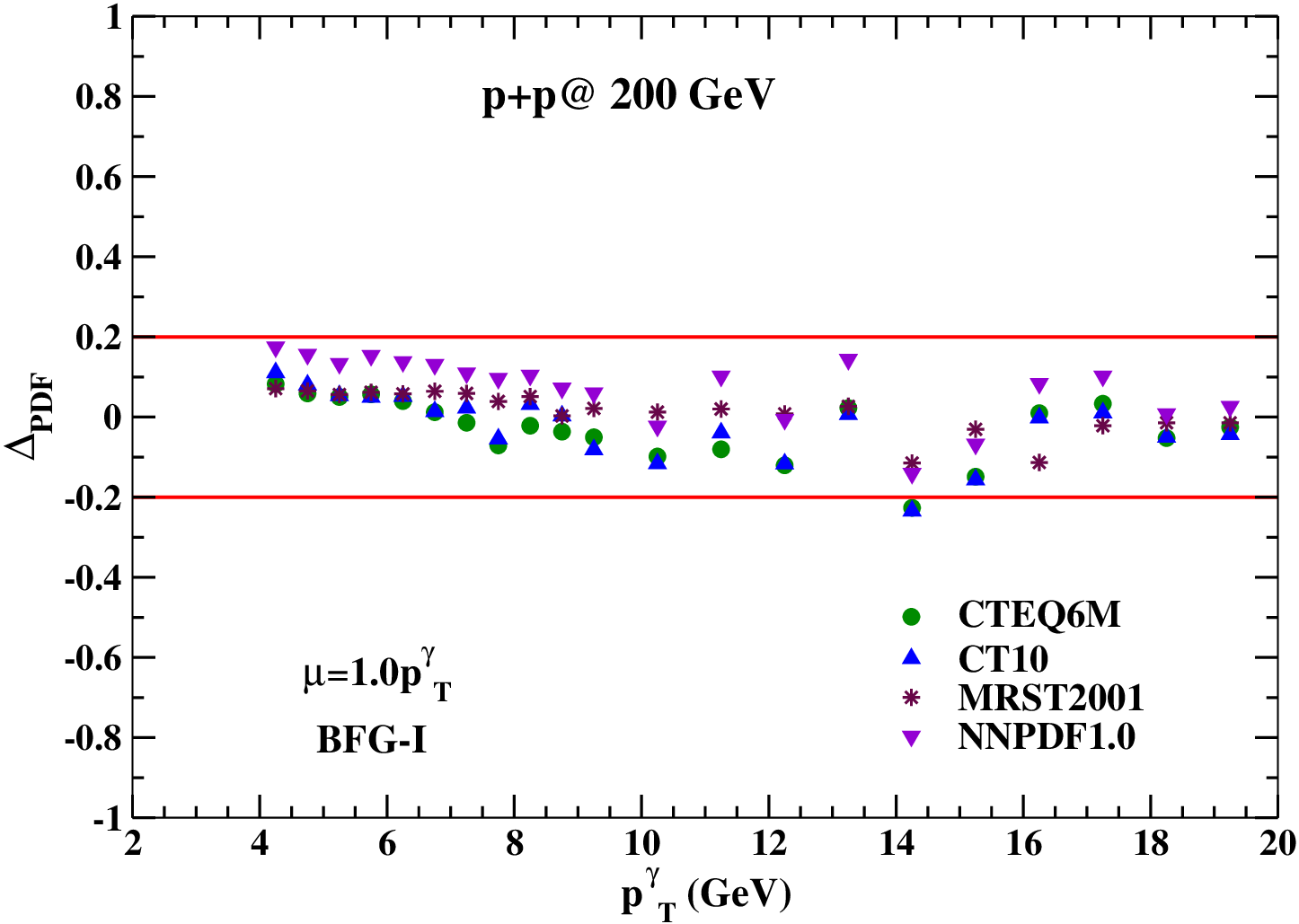}
\includegraphics[width= 6.2 cm, clip= true]{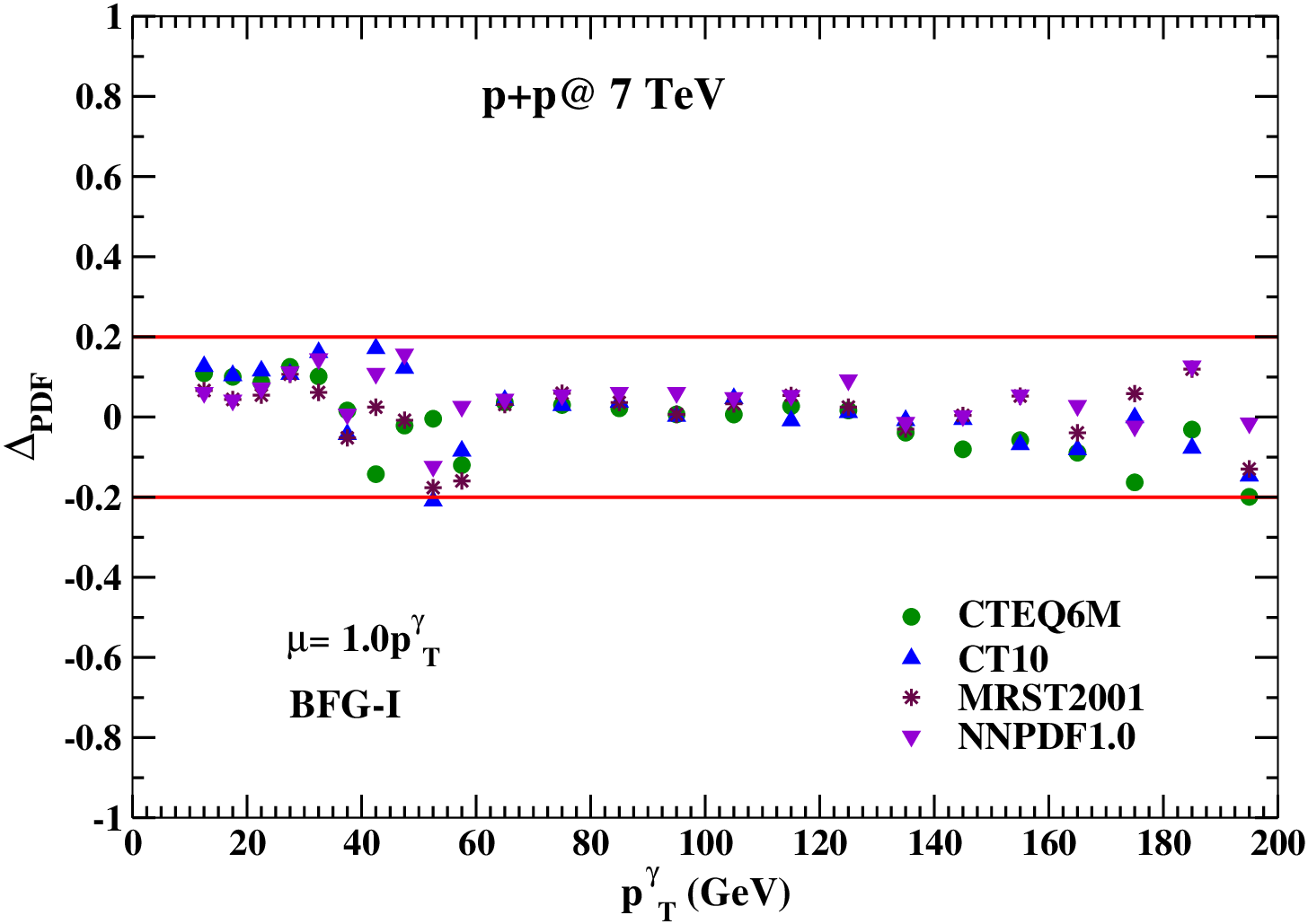}
\end{center}
\caption{(Color Online) The PDF uncertainty in the isolated prompt photon production in p+p collisions related to
the choice of PDF is plotted with the transverse momentum of photon ($\mathrm{p^{\gamma}_T}$) 
at the RHIC (left) and LHC (right) energy.}
\label{fig4}
\end{figure}

Still now we have plotted our results on logerthemic scale which can easily consume a factor of 2--3 difference
among the PDF's. To measure the deviation in the isolated prompt photon yield of different PDF sets, 
we introduce a parameter $\Delta_{\textrm{PDF}} = (\mathrm{Modern-Old})/\mathrm{Old}$. 
The \textquotedblleft$\textrm{Old}$\textquotedblright refers to the isolated prompt photon cross section calculated with CTEQ4M. 
The \textquotedblleft$\textrm{Modern}$\textquotedblright refers to the same calculated with CTEQ6M, CT10, MRST2001 and NNPDF1.0. 
Thus, the CTEQ4M  is chosen as the central value and $\Delta_{\textrm{PDF}}$
measures the deviation as we go towards newer structure functions. The dependence of $\Delta_{PDF}$ with $\mathrm{p^{\gamma}_T}$
at the RHIC and the LHC energies, is displayed in Fig.~\ref{fig4}. 
Apart from the numerical fluctuations, we find all the PDF's show a monotonic variation with $\mathrm{p^{\gamma}_T}$ 
e.g., $\Delta_{\textrm{PDF}}$ of
 CTEQ6M first decreases for $\mathrm{p_T}^\gamma\leq$ 10 GeV at the RHIC and $\mathrm{p^{\gamma}_T}\leq$ 40 GeV at the LHC, 
then increases with $\mathrm{p^{\gamma}_T}$. 
The amount of deviation seen is least for MRST2001 and most for NNPDF1.0. 
It is found that the uncertainty in the isolated photon cross section associated with the choice of PDF's 
is within 20\%  at the RHIC and LHC energy for all $\mathrm{p^{\gamma}_T}$.
Similar work at the Tevatron energy can be found in Ref.~\cite{Ichou} where the comparison is made with three PDF's
and prediction is given for \textrm{p+p} collisions at the top LHC energy ($\sqrt{s}$= 14 TeV). A recent quantitative
study using a particular PDF set and it's replicas at different collider energies has drawn similar conclusion~\cite{Rojo}.

\begin{figure}
\begin{center}
\includegraphics[width= 6.4cm, clip= true]{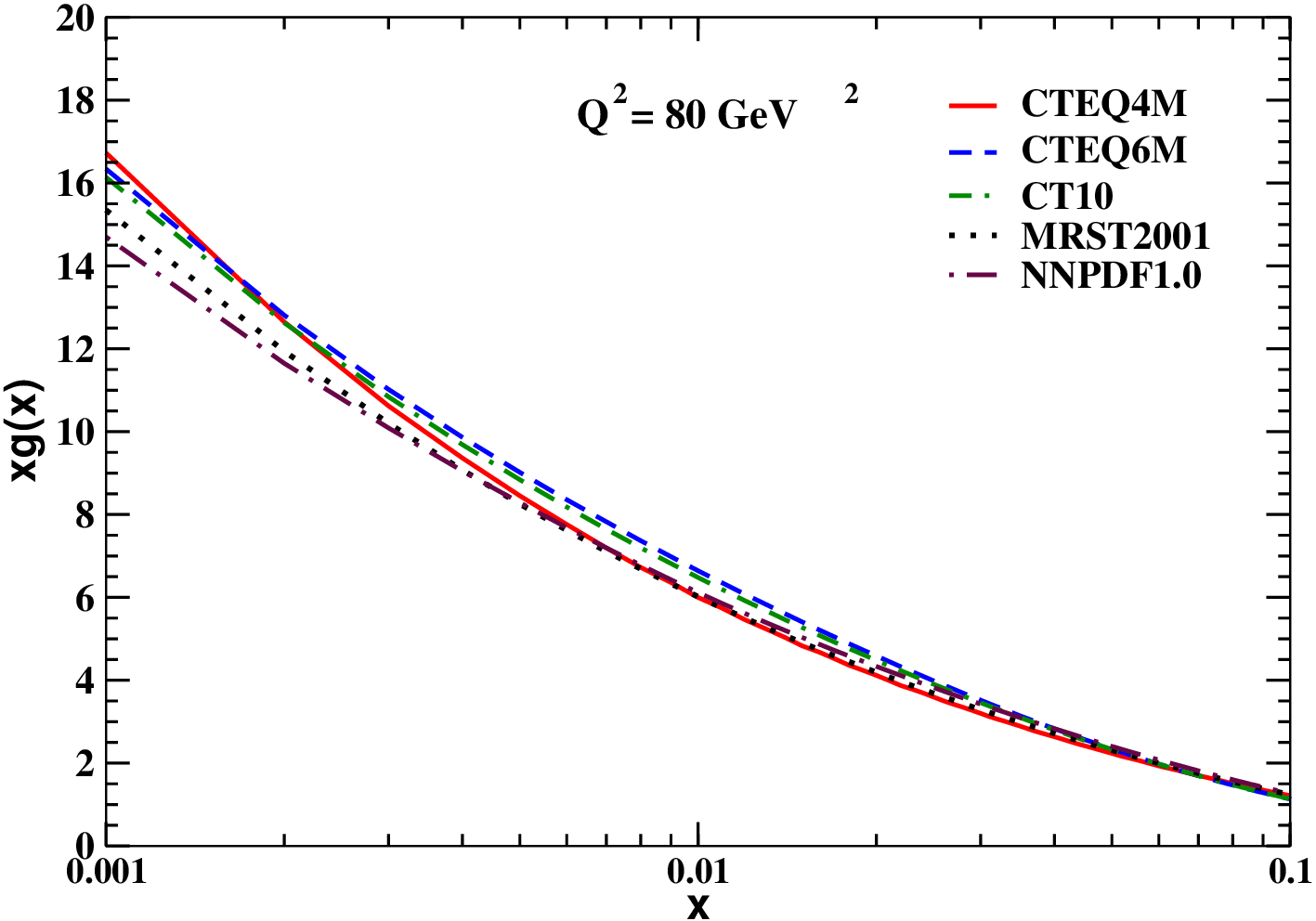}
\end{center}
\caption{(Color Online) The gluon momentum density $x$g($x$,Q$^2$) vs the momentum fraction $x$ for the five PDF's considered
in the present study for a typical value of momentum transfer square Q$^2$= 80 GeV$^2$.}
\label{fig5}
\end{figure}

Finally, we have checked the variation of gluon distribution g($x$,Q$^2$) among the five different parameterizations of PDF
in the kinematic regime, $10^{-3}\leq x\leq10^{-1}$, probed by the present RHIC and LHC photon data. 
In Fig.~\ref{fig5}, we have plotted the momentum density of gluons $x$g($x$,Q$^2$) vs the momentum fraction $x$ 
for a typical value of Q$^2$= 80 GeV$^2$. This shows that the gluon distribution is varying indeed among the five PDF's
considered in this study. We believe that the sensitivity of single prompt photon production to the parton distributions
is taken away due to the integration over the initial parton momenta $x_1$ and $x_2$ in Eq.~\ref{cross}.

However, the double prompt photon production cross section in hadronic collisions, is directly
proportional to the incoming parton momentum density (or the PDF)~\cite{double1,double2}. 
The double prompt photons produced through quark-anti-quark annihilation q\={q}$\rightarrow \gamma\gamma$ or 
gluon-gluon fusion gg$\rightarrow \gamma\gamma$ completely specify parton momentum distribution. The single
bremsstrahlung contribution like; qg$\rightarrow$ q$\gamma\gamma$ to the double photon production is also
sensitive to the PDF. Hence we hope the double prompt photon measurement at the LHC energy 
could better probe the gluon distribution g($x$,Q$^2$) in the proton in comparison with the single prompt photon measurement.

\section{\bf{Summary and Outlook}}
We have analyzed the data of single prompt photon production cross section in \textrm{p+p} collisions at 
the center of mass energies 200 GeV (RHIC) and 7 TeV (LHC) with five different parameterizations of PDF. Our calculation
relies on the available programme \texttt{JETPHOX} which calculates isolated prompt photon production at the 
NLO accuracy. We have shown that the single isolated prompt photon spectra is almost insensitive
to the choice of PDF which is due to the intregation over the initial state parton momemta. 
The present analysis suggests the recent prompt photon measurement at RHIC and LHC can constrain the 
gluon distribution in the proton not more than 20\% of certainty. We expect the differential production cross section of double
prompt photons can determine the gluon distribution more precisely and shall be checked in future studies.

\section*{Acknowledgements}
SD is thankful to Dinesh Srivastava, Rupa Chatterjee for many stimulating discussions and comments on the manuscript.
The author sincerely acknowledges the useful correspondences with Jean-Philippe Guillet regarding \texttt{JETPHOX}.
The author is financially supported by Department of Atomic Energy, India during the course of this work.


\end{document}